# Spectral properties of the one-dimensional Hubbard model

R. Preuss, A. Muramatsu, W. von der Linden, F. F. Assaad and W. Hanke

*Physikalisches Institut, Universität Würzburg, Am Hubland*

*D-97074 Würzburg, Federal Republic of Germany*

## Abstract

The spectral properties of the 1-D Hubbard model are obtained from quantum Monte Carlo simulations using the maximum entropy method. The one-particle excitations are characterized by dispersive cosine-like bands. Velocities for spin- and charge excitations are obtained that lead to a conformal charge $c = 0.98 \pm 0.05$ for the largest system simulated ($N = 84$). An exact sum-rule for the spin-excitations is fulfilled accurately with deviations of at most 10% only around $2k_F$.

Since Anderson [1] proposed that the high-$T_c$ superconductors (HTS) should be considered as Luttinger liquids [2], a great deal of interest was focused on such systems. The best known Luttinger liquids are the Luttinger and the Hubbard model (HM), both in one dimension (1-D). In spite of the fact that both models can be solved exactly [3,4], very little was known about their dynamical properties until recently.

In the case of the Luttinger model, the spectra for collective excitations can be easily obtained by bosonization [3], whereas the one-particle properties were calculated only recently [5,6]. However, since lattice effects are neglected in this model, the results obtained give only the asymptotic behavior for vanishing excitation energy. Being this region the most difficult to be accessed by experiments, further progress is necessary in order to clarify the situation for the HTS.

Lattice effects are contained in the 1-D HM, that has an exact solution by Bethe-Ansatz (BA) [4]. A number of authors succeded in extracting from BA information about spectral properties [7–11], however, many of these results are limited to special situations (e.g. half-filling, $U \to \infty$ and/or one-hole doping, or $\omega = 0$). Only recent progress achieved in the frame of conformal field-theory [12], led to the asymptotic properties of correlation functions irrespective of the coupling constant and doping. Complementary to these achievements, a Landau-Luttinger liquid theory was advanced in order to describe the low-energy properties of the 1-D HM [13]. In spite of the importance of these developments, a general description of spectral properties at finite frequencies is still lacking.

We present in this Letter spectra for one-particle, spin- and charge-density excitations obtained with the maximum entropy method (MEM) [14] from quantum Monte Carlo (QMC) simulations of the 1-D HM. For the first time finite-frequency spectra are presented for system sizes up to 84 sites, allowing, irrespective of doping, for a clear identification of finite-size effects in different quantities . For half-filling, the one-particle spectra show besides the insulating gap, bands that closely follow a cosine-like dispersion. The spectral weight, however, is not evenly spread between the two bands, in contrast to mean-field calculations in the antiferromagnetic state. The magnetic structure form factor shows gapless excitations whereas charge-density excitations show a gap. When the system is doped, a depression appears in the density of states at the Fermi-energy. A single band is observed, that follows very closely a cosine dispersion. In the doped case both spin- and charge-density excitations are gapless, with a "holon"-velocity larger than the "spinon" one, such that charge-spin separation is manifest. The spinon and holon velocities obtained lead to a value of the conformal charge of $c = 0.98 \pm 0.05$ for $N = 84$ sites, in very good agreement with the exact value $c = 1$ [12]. Finally, it is shown, that the spectrum for magnetic excitations fulfills excellently a frequency sum rule [15] for each $k$-point except for those close to $2k_F$, where departures of at most 10% are observed.

We consider the 1-D HM described by the following Hamiltonian:

$$H = -t \sum_{i,\sigma} \left( c^\dagger_{i+1,\sigma} c_{i,\sigma} + h.c. \right) + U \sum_i n_{i\uparrow} n_{i\downarrow} , \qquad (0.1)$$

where $c^{(\dagger)}_{i,\sigma}$ are annihilation (creation) operators for an electron at site $i$ with spin $\sigma$, and $n_{i\sigma} = c^\dagger_{i,\sigma} c_{i,\sigma}$. Systems with a number of sites $N$ ranging from $N = 12$ to $N = 84$ with periodic boundary conditions were simulated for inverse temperatures $\beta = 1/k_B T$ up to $\beta = 20/t$ and interaction strength $U = 4t$. The simulations were performed with the grand canonical algorithm [16], where the smallest values of $\Delta\tau$, i.e. the time-slice, was between 0.1 and 0.125. The analytic continuation of the data to real frequencies was performed with the MEM [14], where the only "prior knowledge" used was the positivity of the spectral functions. We have chosen an uninformative default model $m_i = \varepsilon$, where $\varepsilon$ is a small quantitiy that merely suppresses noise in regions of insufficient information. The regularization parameter has been determined selfconsistently by classical MEM. For further technical details we refer to previous applications of the MEM to the single impurity Anderson model [17], the one- [18] and two- [19] dimensional Heisenberg antiferromagnet.

We first compare the density of states $D(\omega)$ of a ring with 12 sites between data from QMC simulations and exact diagonalization (ED), both for half-filling (Fig. 1a) and for the doped case (Fig. 1b) with $<n> = 0.833 \simeq 5/6$. Although the MEM is not able to resolve the rich structure obtained in ED, there is a good agreement in the shape and distribution of weight in the half-filled case. In the doped case, the MEM can reproduce well the structures close to the Fermi-energy but has the general tendency of shifting weight to lower energies as one goes to higher energies due to the Laplace-transformation kernel. In particular, the pseudogap observed in ED that separates those states, whose weight was transferred upon doping from the upper Hubbard band (UHB) to the top of the lower Hubbard band (LHB) [20], from the remaining of the UHB, does not appear in the MEM data.

In the following the data obtained for our largest system ($N = 84$), well beyond the capability of ED are shown [21]. We discuss first the spectral properties at half-filling. Figure 2a shows $D(\omega)$ and the spectral function $A(k,\omega)$ is shown in Fig. 2b, where a dispersion can be clearly seen. Figure 2c shows the location of the maxima of $A(k,\omega)$ with the errors assigned by the MEM. These errors give the uncertainty in the location of those maxima. The full curve is just a cosine function with a band-width that fits the QMC data. Remarkably, the dispersion obtained closely follows the two bands that would be obtained in a mean-field type of calculation, where antiferromagnetic order is assumed. However, the spectral weight is shared quite differently, since although most of the weight of the LHB appears for $k \lesssim \frac{\pi}{2}$, and the weight of the UHB is mostly concentrated in the region $k \gtrsim \frac{\pi}{2}$, around $k = \frac{\pi}{2}$ weight is splitted between the two bands.

Figure 3 shows the imaginary part of the spin-susceptibility $\chi_S(k,\omega)$ and the dispersion extracted from the maxima of Fig. 3a. From the slope of the dispersion around $k = 0$, we calculate a spinon-velocity $v_s/t = 1.23 \pm 0.11$. The exact value extracted from BA [7] $v_s^{BA} = 2tI_1(2\pi/U)/I_0(2\pi/U)$, where $I_0$ and $I_1$ are modified Bessel functions, is for $U/t = 4$, $v_s^{BA}/t = 1.2263$ [22], that compares remarkably well with our value. The full line in Fig. 3b corresponds to the dispersion extracted from Bethe Ansatz [7] showing that the agreement with the exact results in the thermodynamic limit is not limited to the low-energy limit but extends to finite frequencies. For reasons of space, the equivalent quantities to those in Fig. 3 but now for charge-density excitations, are not displayed. We observe a rather broad structure with most of the weight around $U$, with a gap for $k \to 0$ that agrees within the errors with the one obtained in the one-particle spectrum.

Now we consider the doped case, where for definitness we have chosen a hole doping $\delta \simeq \frac{1}{6}$. Figure 4 shows $D(\omega)$, $A(k,\omega)$, and the dispersion of the maxima in $A(k,\omega)$. A depression in $D(\omega)$ is observed at the Fermi-energy $\epsilon_F$, that we assign to the fact that a Luttinger liquid has a density of states $D(\omega) \sim |\omega|^\alpha$, with $\alpha \simeq 0.038$ for $U/t = 4$ [23]. The vanishing of $D(\omega)$ at the Fermi energy cannot be seen in the simulation, since due to the small power $\alpha$, an unrealistically high resolution would be necessary. Again, the "band-structure" obtained from $A(k,\omega)$ can be very well described by a cosine band within the error bars. It should be stressed, that $A(k,\omega)$ shows a peak that sharpens as one approaches $\epsilon_F$ for all the sizes simulated ($N = 12, 22, 26, 36, 50, 60, 70$ and $84$). This seems to contradict the common lore of a Luttinger liquid having vanishing spectral weight at $\epsilon_F$. However, Sorella and Parola have shown [11] in the limit $U \to \infty$ and for low density, that the quasiparticle weight $z(k_F)$ vanishes as $N^{-1/8}$, where the power coincides with $\alpha$, the exponent characterizing the momentum distribution function around $k_F$ and $D(\omega)$ around $\epsilon_F$ [23]. For the one-hole case, on the other hand, they obtained $z(k_F) \sim N^{-1/2}$. Therefore, we should expect for $U = 4$ an even lower exponent [24], and hence, an utopically large system is needed to decide whether the system is a Fermi-liquid or not, on the basis of $A(k,\omega)$. At U=4 we still do not observe significant weight in $A(k,\omega)$ due to the UHB. However as was discussed in connection with ED, this can be due to the shifting of weight to lower frequencies by the MEM.

The spectra for spin-excitations are shown in Fig. 5. Again, a linear dispersion appears for low energies around $k = 0$ (inset of Fig. 5). Unfortunately the same resolution is not obtained around $2k_F = 5\pi/6$. However, these data can be combined with the corresponding ones for charge-density excitations (Fig. 6). With the values of spinon and holon velocities extracted from the dispersion curves, it is possible to obtain the conformal charge of the HM, that is known exactly to be $c = 1$. Such a relationship stems from the conformal invariance of the model and has the following form [12]:

$$\frac{E_0(N)}{N} - \epsilon_0 = -\frac{\pi}{6N^2}(v_s + v_c)c + \mathcal{O}\left(\frac{1}{N^2(\ln N)^3}\right), \qquad (0.2)$$

where $E_0(N)$ is the ground-state energy of the system with $N$ sites, $\epsilon_0$ is the ground-state energy per site in the thermodynamic limit, $v_s$ and $v_c$ are the spinon and holon velocities, respectively, and $c$ is the conformal charge of the model. The value obtained for $N = 84$ is $c = 0.98 \pm 0.05$, where $E_0$ and $\epsilon_0$ were determined from Bethe-Ansatz [4]. Such a stringent test shows that QMC together with the MEM are able to give reliable real frequency data for low-lying excitations of the HM. It should be stressed, that the insets in Figs. 5 and 6 make manifest charge-spin separation in the model, since $v_c > v_s$. This is clearly seen in all system sizes simulated. Finally, we consider an exact sum-rule for the first frequency-moment of $\text{Im}\chi_S(k,\omega)$ [15]:

$$\int_{-\infty}^{\infty} \frac{d\omega}{2\pi} \omega \, \text{Im}\chi_S(k,\omega) = - < [[H, S^z(k)], S^z(-k)] >$$
$$= -2t(1 - cosk) < H_{kin} >, \qquad (0.3)$$

where $S^z = \frac{1}{2}(n_\uparrow - n_\downarrow)$ and $< H_{kin} >$ is the expectation value of the kinetic energy, that can be accurately calculated by QMC. We find [21] that the sum rule is fulfilled accurately over most of the Brillouin-zone (deviations of less than 1%), with the exception of $k \gtrsim 2k_F$, where deviations ($\sim 10\%$) are obtained. They are probably due to a broad continuum similar to the one present in the 1-D Heisenberg antiferromagnet around $k = \pi$ [13,26]. This result together with the conformal charge obtained, demonstrate the degree of reliability of the numerical data from low to intermediate frequencies and for very large systems that are well beyond the capability of other methods like ED.

Summarizing, we have presented real-frequency spectra for one-particle, spin- and charge-density excitations in the 1-D Hubbard model both in the insulating and in the metallic phases. The one-particle excitations show a band-like dispersion that can be accurately fitted by a cosine function. At half-filling, spin-excitations are gapless, whereas charge excitations show in this case a gap. The dispersions obtained agree with results from BA within the error bars. In the doped case, charge-spin separation is obtained, where the spinon and holon velocities excellently agree with the exact value for the conformal charge of the Hubbard model. The quality of the spectral data at intermediate frequencies is checked by an exact sum-rule, showing that only around $2k_F$, a departure of around 10% is present.

We would like to thank T. Pruschke for instructive discussions. We are grateful to the Bavarian "FORSUPRA" program on high $T_c$ research for financial support. The calculations were performed at the Cray YMP of the HLRZ Jülich and at LRZ München under a cooperation program with Cray Research Inc. We thank the above institutions for their support.

FIGURES

FIG. 1. A comparison between QMC/MEM (lines with dots) and ED (full line) of the density of states for $N = 12$ sites: (a) Half-filling; (b) $<n>= 5/6$ ($\beta = 16$, $U = 4$).

FIG. 2. One-particle excitation for 84 sites at half-filling: (a) Density of states; (b) $A(k,\omega)$; (c) maxima of $A(k,\omega)$, the solid line is a cosine function adjusted to the bandwith. ($\beta = 20$, $U = 4$)

FIG. 3. Spin-excitations at half-filling: (a) $\mathrm{Im}\chi_S(k,\omega)$; (b) maxima of $\mathrm{Im}\chi_S(k,\omega)$, the dotted line gives the slope at $k = 0$ ($v_s/t = 1.23 \pm 0.11$), and the full line stems from Bethe-Ansatz.

FIG. 4. The same as Fig. 2, but now for $<n>= 5/6$.

FIG. 5. $\mathrm{Im}\chi_S(k,\omega)$ for $<n>= 5/6$. The inset shows the linear part of the dispersion around k=0. The slope gives $v_s/t = 1.33 \pm 0.09$.

FIG. 6. The same as Fig. 5 but for the charge susceptibility ($v_c/t = 2.02 \pm 0.14$).

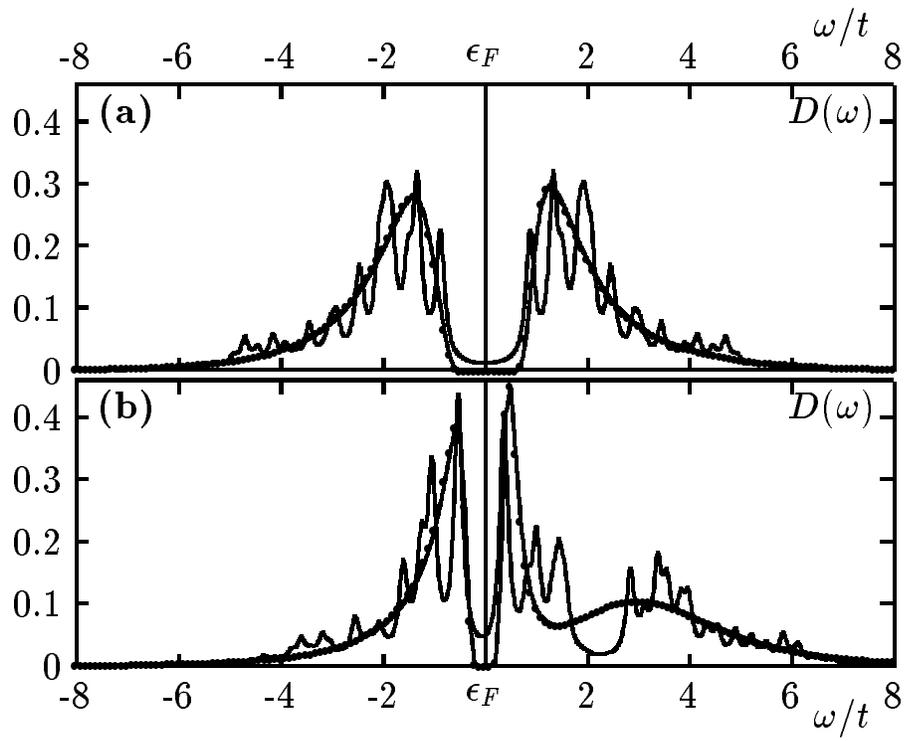

Fig. 1

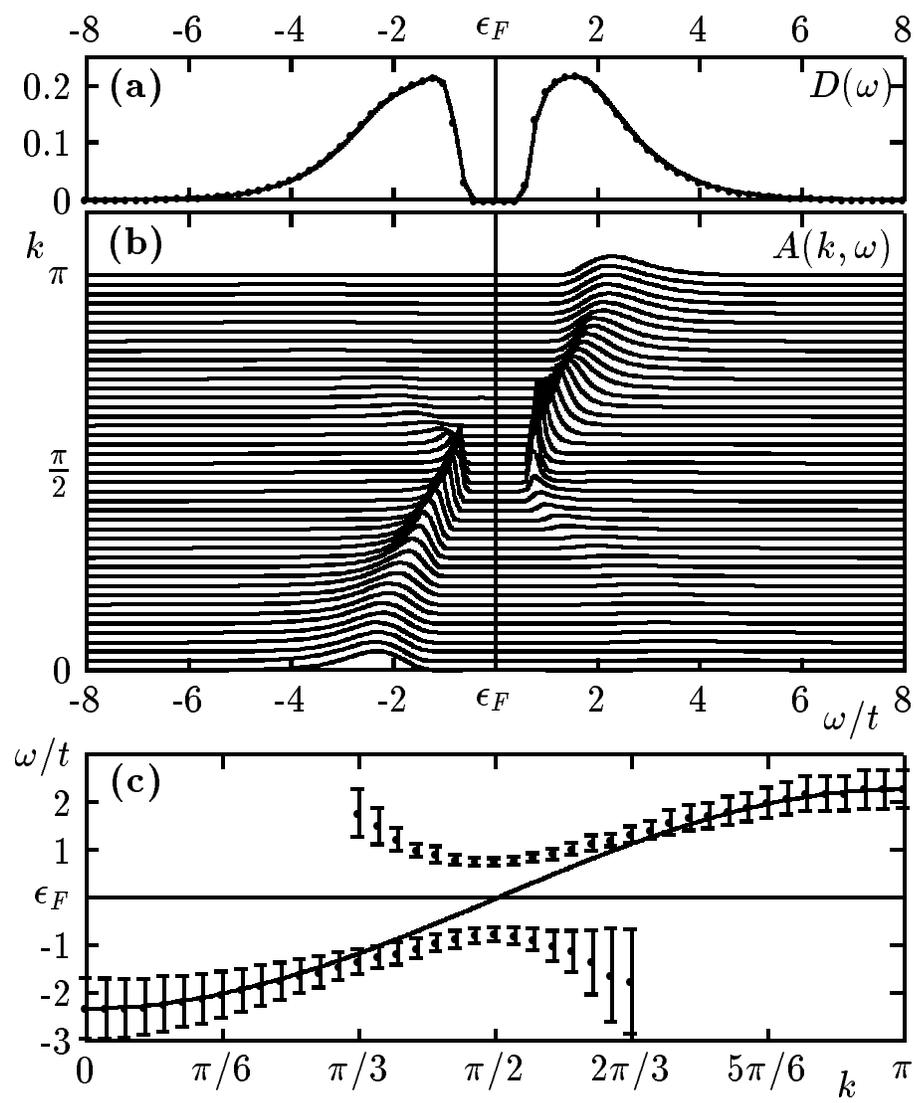

Fig. 2

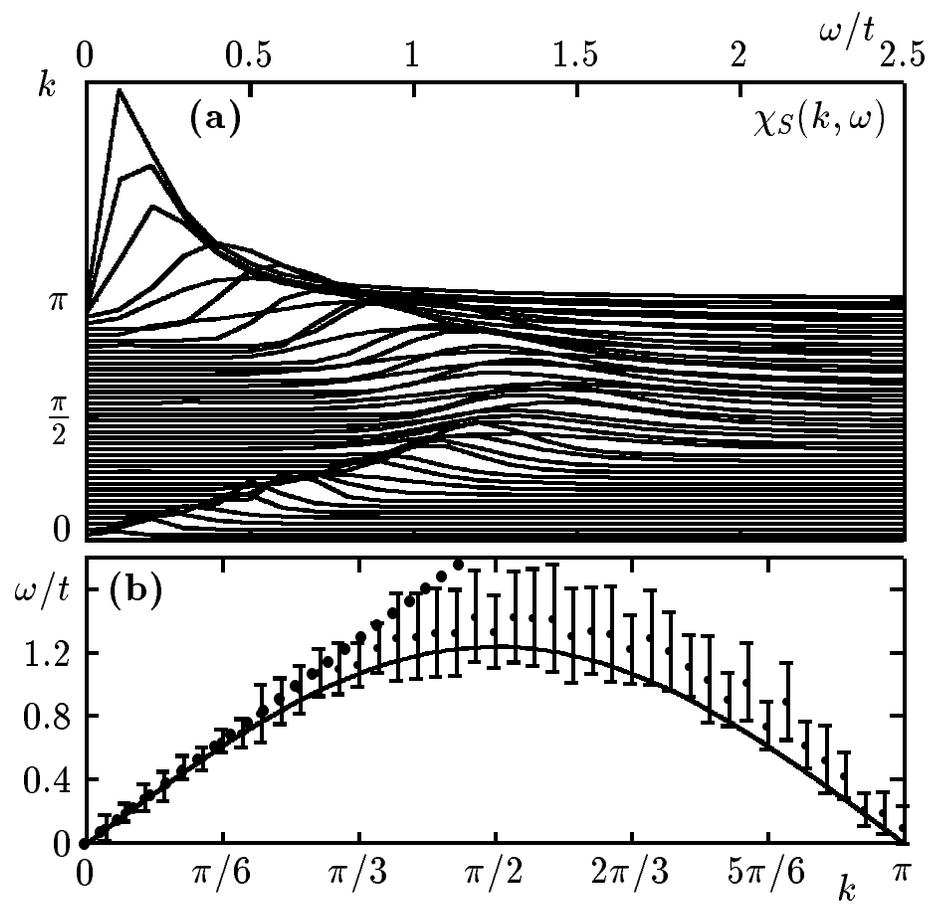

Fig. 3

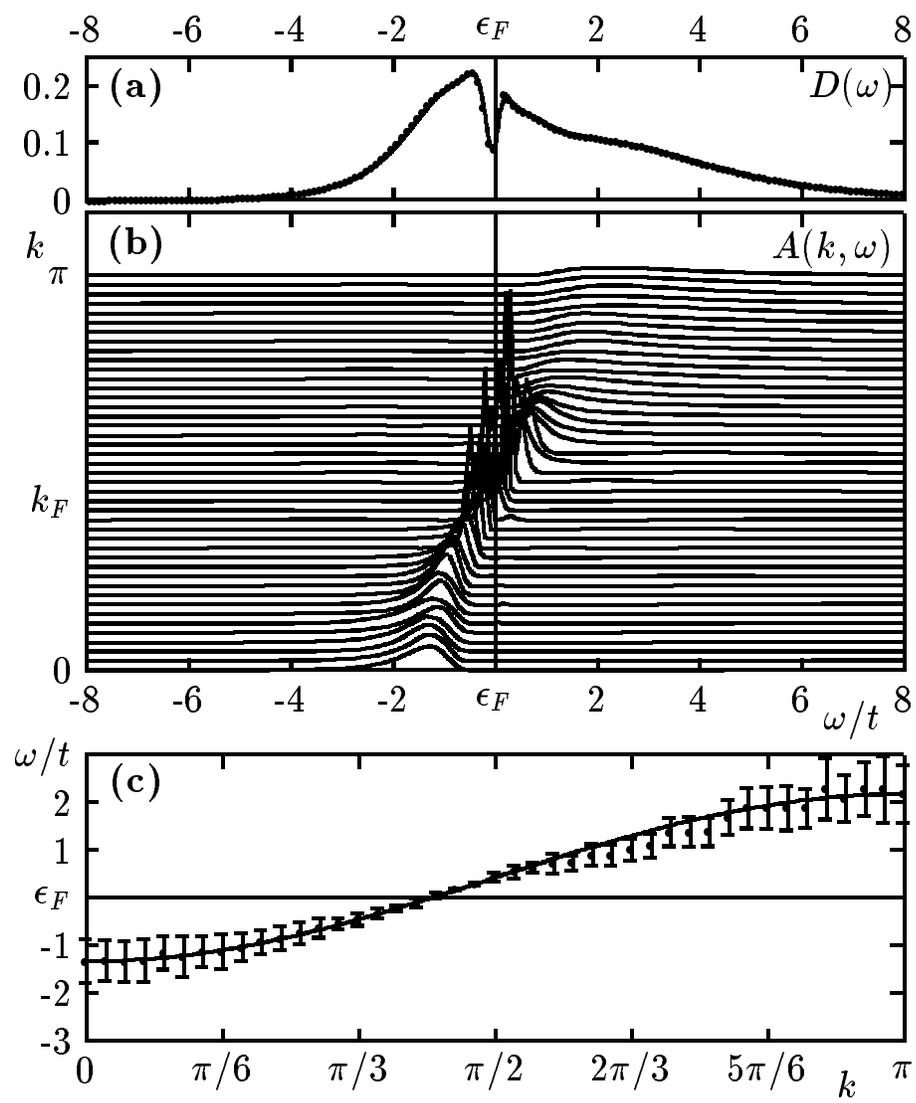

Fig. 4

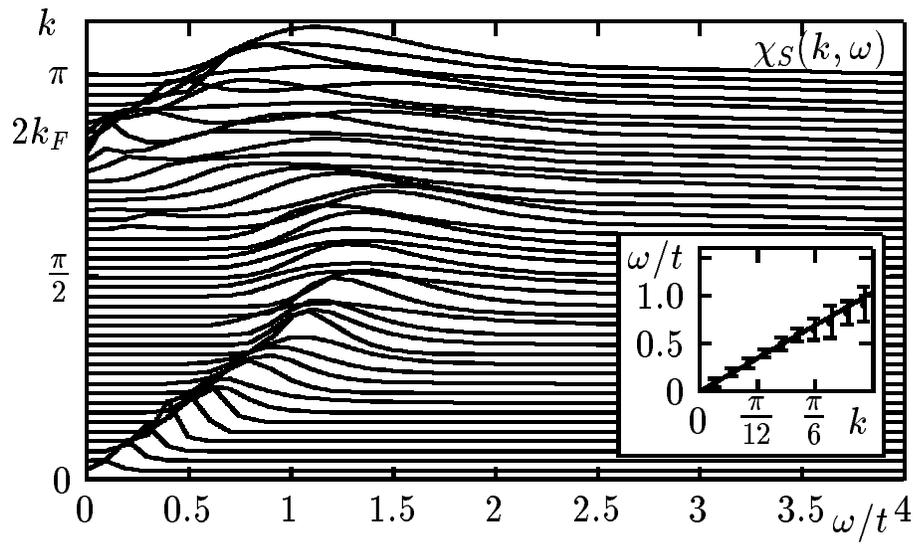

Fig. 5

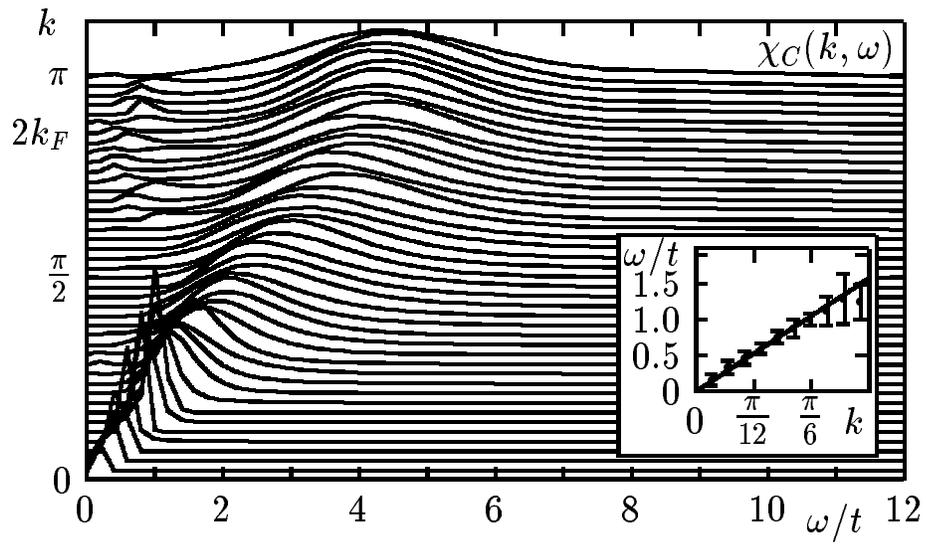

Fig. 6